\documentclass[12pt]{iopart}
\usepackage{iopams,graphicx,bm,color}  
\newcommand\MMM[3]{{#3} {\it J. Magn. and Magn. Mater.} {\bf {#1}} {#2}}

\newcommand\PRB[3]{{#3} {\it Phys. Rev. B} {\bf {#1}} {#2}}
\newcommand\PCM[3]{{#3} {\it J. Phys.: Cond. Matt.} {\bf {#1}} {#2}}
\newcommand\PhysB[3]{{#3} Physica B {\bf {#1}} {#2}}
\newcommand\PPRL[3]{{#3} {\it Phys. Rev. Lett.} {\bf {#1}} {#2}}
\newcommand\JPCS[3]{{#3} {\it J. Phys.: Conf. Series} {\bf {#1}} {#2}}
\newcommand\seo{$\rm SrEr_2O_4$}
\newcommand\sdo{$\rm SrDy_2O_4$}
\newcommand\sho{$\rm SrHo_2O_4$}
\newcommand\syo{$\rm SrYb_2O_4$}
\newcommand\slo{Sr$Ln_2$O$_4$}

\begin{document}
\title[Magnetic field induced ordering in SrDy$_2$O$_4$]{Magnetic field induced ordering in SrDy$_2$O$_4$}
\author{T H Cheffings, M R Lees,  G Balakrishnan and O A Petrenko}
\address{University of Warwick, Department of Physics, Coventry, CV4~7AL, UK}
\ead{o.petrenko@warwick.ac.uk}
\begin{abstract}
Heat capacity measurements were used to investigate the magnetic ordering processes in single crystal samples of \sdo\ in a magnetic field applied along the [010] and [001] directions.
In zero field this compound appears to be magnetically disordered down to at least~0.39~K.
A magnetic field applied along the [010] direction induces a very sharp transition at 20~kOe, seen as a strong peak in the heat capacity versus field, $C(H)$ curves, while for $H \parallel [001]$, the magnetisation process is accompanied by the development of only broad features in the $C(H)$ curves.
The process of field induced ordering in \sdo\ appears to be rather remarkable even in the context of the unusual phase transitions observed in other geometrically frustrated magnetic systems consisting of hexagons and triangles.
\end{abstract}
\date{\today}
\section{Introduction}
The effects of geometrical frustration on the magnetic properties of materials are rich and often unexpected~\cite{books}.
One of the most spectacular effects is the absence of a long-range order (LRO) down to the lowest temperatures despite the presence of relatively strong magnetic interactions.
In some cases magnetic order can be induced by an applied magnetic field, but the number of materials where such an effect has been detected is very limited~\cite{caution}, with perhaps most notable examples found among the garnets~\cite{GGG} and the pyrochlores~\cite{pyros}.
\sdo\ reported here constitutes an interesting addition to this short list.

\begin{figure}
\includegraphics[width={0.5\columnwidth}]{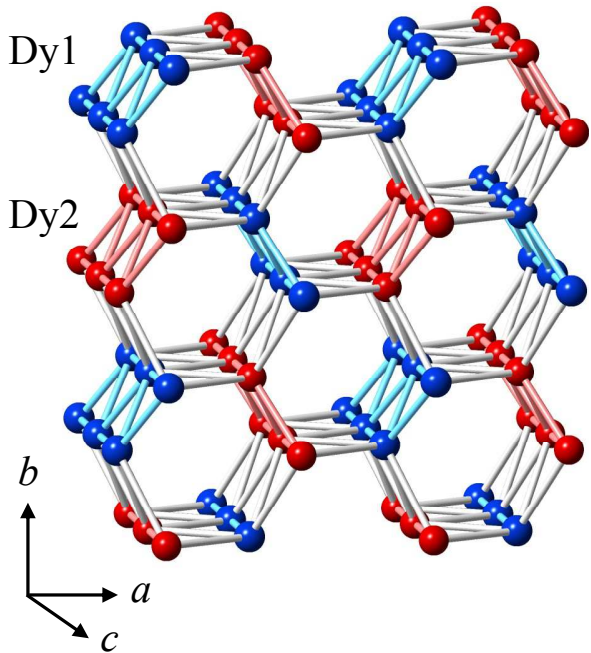}
\caption{(Colour online)	Magnetic sublattice of \sdo, with the two crystallographically inequivalent positions of Dy ions shown in different colours.
					When viewed along the c axis, honeycombs of Dy$^{3+}$ ions are visible.
					Zigzag ladders running along the c axis connect the honeycomb layers and give rise to geometric frustration.}
\label{Fig1}
\end{figure}

\sdo\ belongs to the family of rare-earth strontium oxides, \slo, which crystallise in the form of calcium ferrite, with the space group $Pnam$.
The crystal structure of these materials (see Fig.~\ref{Fig1}) can be viewed as a network of linked hexagons and triangles~\cite{Karunadasa_2005}, and their magnetic properties have been studied within the context of geometrically frustrated magnetism.
Several members of the family have been the subject of recent investigations~\cite{Ghosh_2011,Quintero_2012,Petrenko_2008,Hayes_2011,Young_2012}.

\syo\ is reported to order magnetically at $T_N=0.9$~K into a noncollinear structure~\cite{Quintero_2012}, with a significantly reduced ordered spin moment on the magnetic Yb$^{3+}$ sites compared to the full ionic moment of Yb$^{3+}$.
The magnetic $H-T$ phase diagram of \syo\ consists of a complicated series of states, which are attributed to the competition of the various magnetic interactions with each other as well as with the relatively strong single-ion anisotropy~\cite{Quintero_2012}.

With the help of powder neutron diffraction (PND) data, it has been established~\cite{Petrenko_2008} that \seo\ orders magnetically at $T_N=0.75$~K.
However, single-crystal neutron diffraction~\cite{Hayes_2011} has revealed two distinct components to the magnetic ordering in this compound: one component is a LRO  {\bf k}~=~0 structure which appears below $T_N=0.75$~K, and the other component is a short-range incommensurate structure which is responsible for the appearance of a strong diffuse scattering signal.
The diffuse scattering in this compound is observed to form undulating planes of intensity at the positions $(h,k,\frac{1}{2}+\delta)$ and $(h,k,\frac{3}{2}-\delta)$, with the incommensuration parameter $\delta$ varying from 0.2 to 0.6, depending on the temperature.
The partially ordered component does not undergo a pronounced phase transition at any temperature down to 60~mK~\cite{Hayes_2011}.

PND data obtained on \sho\ \cite{Young_2012} seem to indicate a magnetic structure very similar to the one observed in \seo, in that a LRO {\bf k}~=~0 structure coexists with a short-range diffuse component.
However, more precise single-crystal neutron diffraction measurements~\cite{Young_2013} suggest that even the {\bf k}~=~0 component appearing below $T_N=0.68$~K is short-range in \sho.
 It is also observed~\cite{Young_2013} that the planes of scattering intensity formed in reciprocal space by the diffuse component are not as undulated in their position as in \seo, which suggests that \sho\ should be regarded as a collection of almost perfect one-dimensional spin chains.
 Surprisingly, despite the absence of the LRO, the $T_N$ of 0.68~K is still marked by a cusp in the susceptibility as well as by a noticeable difference between the field-cooled and the zero-field-cooled data below this temperature~\cite{Hayes_2012}.

PND data for \sdo\ (SDO) show no signs of any long-range magnetic order down to 20~mK, as the scattering pattern in zero field is dominated by broad diffuse scattering peaks~\cite{Petrenko_unpublished}.
Previous powder~\cite{Karunadasa_2005} and single-crystal~\cite{Hayes_2012} susceptibility measurements have shown that the magnetic moment on the Dy$^{3+}$ sites is rather large, $\gtrsim 10 \; \mu_B$, approaching the maximum expected value of 10.63~$\mu_B$.
With a reported~\cite{Karunadasa_2005} Curie-Weiss temperature of -23~K for SDO compared to -17~K for \sho\ and -13.5~K for \seo, the lack of magnetic ordering is rather surprising.
In this paper we report on the magnetic ordering processes in SDO induced by an applied magnetic field.
The ordering is investigated by measuring the temperature and field dependence of the heat capacity of single crystal samples for $H \parallel [010]$ and $H \parallel [001]$.
\section{Experimental procedures}
Single crystal samples of \sdo\ were prepared as described previously~\cite{Balakrishnan_2009}.
The principal axes of the samples were determined using the Laue x-ray diffraction technique; the crystals were aligned to within an accuracy of 3 degrees.

Specific heat measurements were performed in the temperature range 0.39 to 5.0~K, in fields of up to 50~kOe using a Quantum Design Physical Property Measurement System (PPMS) calorimeter equipped with a $^3$He option.
We have measured the specific heat as a function of temperature in a constant magnetic field and as a function of the applied field at a constant temperature.
For field-scans, the temperature stability was better than~5~mK. 
The measurements were carried out on small (typically less than 1~mg) platelike samples for which the demagnetising factors were small.
The measurements were restricted to the $H \parallel [010]$ and $H \parallel [001]$ directions, as applying a magnetic field along $a$, the ``hard-axis"  for magnetisation~\cite{Hayes_2012}, would cause a significant torque on the heat capacity sample platform.
For the same reason, the measurements for $H \parallel [001]$ are limited to a maximum field of 30~kOe, as above this field the difference between $M_{H \parallel [001]}$ and $M_{H \parallel [010]}$ becomes substantial~\cite{Hayes_2012}.

From previous measurements~\cite{Petrenko_2008} of the heat capacity of SrLu$_2$O$_4$ and SrY$_2$O$_4$, two nonmagnetic compounds isostructural with SDO, it may be concluded that the lattice contribution to the specific heat is negligibly small compared to the magnetic contribution for all temperatures below 5~K, therefore all the heat capacity data shown below should be considered to be magnetic in origin.
\section{Experimental results}
\subsection{Zero-field data}
\begin{figure}
\includegraphics[width={0.5\columnwidth}]{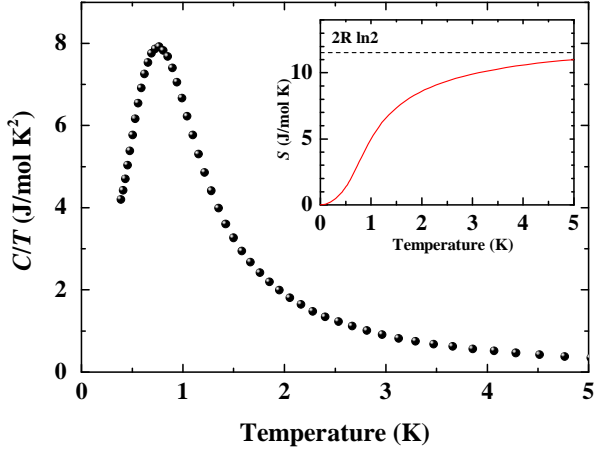}
\caption{Temperature dependence of the specific heat divided by temperature of \sdo\ in zero field.
	The inset shows the temperature dependence of the entropy, $S$ (solid line), calculated as the area under the $C(T)/T$ curve which has been extended linearly down to $T=0$~K.
	The dashed line indicates the position of 2R$\ln(2)$, which corresponds to the magnetic contribution for a system with an effective $s=1/2$.}
\label{Fig2}
\end{figure}
The temperature dependence of the heat capacity divided by temperature of SDO is shown in Fig.~\ref{Fig2}.
The $C(T)/T$ curve shows a very broad maximum at 0.77~K and a nearly linear temperature dependence below this peak, implying a quadratic, $T^2$, dependence for $C(T)$ at low temperatures.
There are no sharp features in the heat capacity curve which could be attributed to a phase transition to a magnetically ordered state. 
The inset shows the temperature dependence of magnetic entropy obtained from the $C(T)/T$ curve extrapolated to $T=0$, as well as the entropy value expected for an effective spin $1/2$ system.
\subsection{$H \parallel [010]$}
\begin{figure}
\includegraphics[width={0.5\columnwidth}]{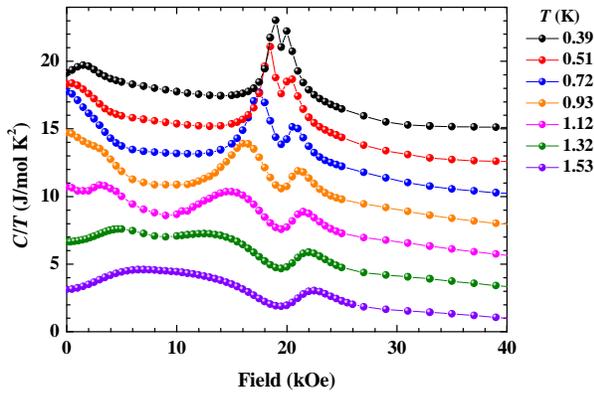}
\caption{(Colour online)	Field dependence of the heat capacity divided by temperature, $C(T)/T$, of single crystal \sdo\ measured at different temperatures for $H \parallel [010]$.
					The curves are consecutively offset by 2.5 J/(mol K$^2$) for clarity.}
\label{Fig3}
\end{figure}
The field dependence of the heat capacity divided by temperature of SDO measured at different temperatures for $H \parallel [010]$ is shown in Fig.~\ref{Fig3}.
At the lowest temperature of 0.39~K, a sharp double peak at about 20~kOe dominates the $C(H)$ curve.
On warming the sample, the peaks in the $C(H)$ curve become more rounded while the field separation between them increases.
Above 1~K, the peaks (especially the one at lower-field) are rather broad and separated by approximately 6~kOe, while by 1.5~K the lower-field peak is too broad and too weak to be clearly distinguishable.
Apart from these two peaks, a low-field ($< 4$~kOe) feature with a pronounced temperature dependence is also clearly visible on the $C(H)$ curves.

\begin{figure}
\includegraphics[width={0.5\columnwidth}]{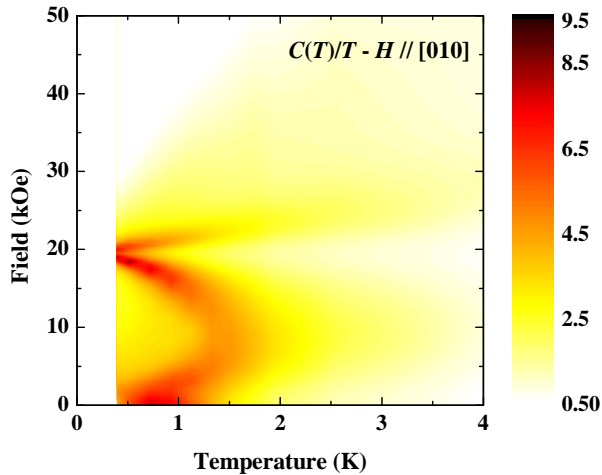}
\caption{(Colour online)	Magnetic $H-T$ phase diagram of \sdo\ for $H \parallel [010]$ obtained from the heat capacity measurements.
					Colour represents the heat capacity divided by temperature in the units of J/(mol K$^2$).}
\label{Fig4}
\end{figure}

The $C(H)/T$ curves shown in Fig.~\ref{Fig3} have been combined with similar field scans performed at higher temperatures to form the magnetic $H-T$ phase diagram shown in Fig.~\ref{Fig4}, where the value of the heat capacity divided by temperature is represented by the colour scale shown on the right of the figure. 
\subsection{$H \parallel [001]$}
\begin{figure}
\includegraphics[width={0.5\columnwidth}]{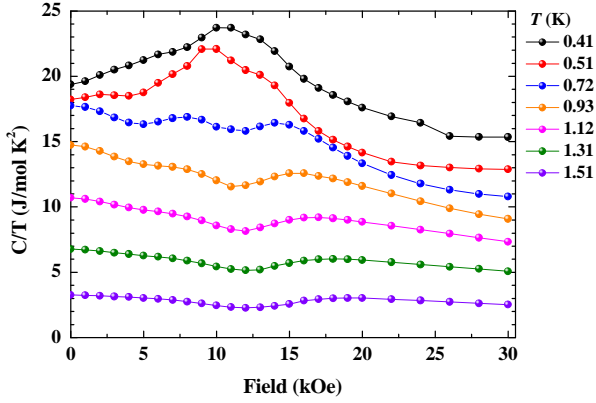}
\caption{(Colour online)	Field dependence of the heat capacity divided by temperature, $C(T)/T$, of single crystal \sdo\  measured at different temperatures for $H \parallel [001]$.
					The curves are consecutively offset by~2.5~J/(mol K$^2$) for clarity.}
\label{Fig5}
\end{figure}

The field dependence of the heat capacity divided by temperature of SDO measured at different temperatures for $H \parallel [001]$ is shown in Fig.~\ref{Fig5}.
In contrast to the case for a field applied along the $b$ axis, the application of a magnetic field along the $c$ direction does not seem to result in any features in the $C(H)$ curves sharp enough to be indicative of a phase transition.
At the lowest temperature, a broad peak in $C(H)$ is present at around~11~kOe.
On warming the sample this peak initially increases in intensity, but then decreases rapidly and splits into two very broad peaks.
By~0.93~K the same field of 11~kOe corresponds to a local {\it minimum}, rather than a {\it maximum} in the heat capacity.
This minimum remains visible (albeit less pronounced) at higher temperatures and seems to shift to slightly higher fields.

\begin{figure}
\includegraphics[width={0.5\columnwidth}]{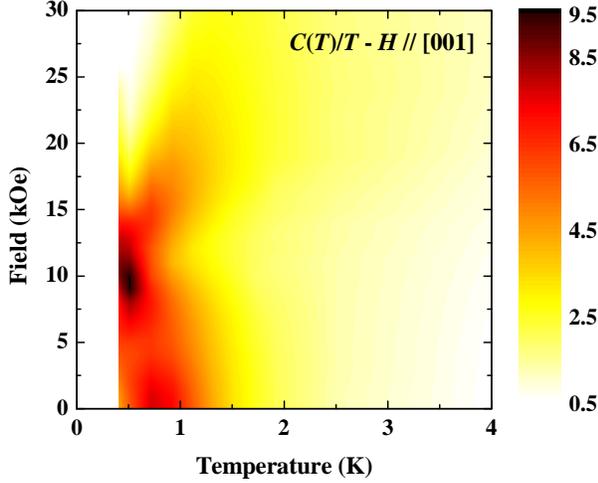}
\caption{(Colour online)	Magnetic $H-T$ phase diagram of \sdo\ for $H \parallel [001]$ as obtained from the heat capacity measurements.
					Colour represents the heat capacity divided by temperature in the units of J/(mol K$^2$). }
\label{Fig6}
\end{figure}

\begin{figure}
\includegraphics[width={0.5\columnwidth}]{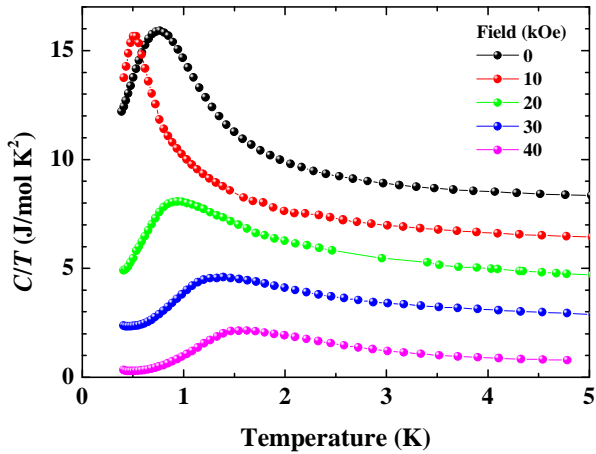}
\caption{(Colour online)	Temperature dependence of the heat capacity divided by temperature, $C(T)/T$, of single crystal \sdo\ measured in different applied fields for $H \parallel [001]$.
					The curves are consecutively offset by~2.0~J/(mol K$^2$) for clarity.}
\label{Fig7}
\end{figure}

The $C(H)/T$ curves shown in Fig.~\ref{Fig5} are combined with similar field scans performed at higher temperatures to form the magnetic $H-T$ phase diagram shown in Fig.~\ref{Fig6}, where the variation in the heat capacity divided by temperature is represented by the colour scale.
Since for this direction of applied field there are no sharp features in the $C(H)/T$ curves, we have performed additional measurements of the temperature dependence of the specific heat in constant fields; the corresponding data are shown in Fig.~\ref{Fig7}.
Again, no sharp peaks are observed in the $C(T)$ curves.
A broad peak centred at 0.77~K in zero field shifts to 0.5~K and becomes more intense and well defined in an applied field of 10~kOe.
With the further increases in field the peak shifts to higher temperatures and becomes significantly less intense.  

\section{Discussion and Conclusions}
The zero-field heat capacity data shown in Fig.~\ref{Fig2} implies the absence of long-range magnetic order in SDO down to at least~0.39~K in full agreement with our PND data~\cite{Petrenko_unpublished} which
show only broad diffuse scattering peaks down to 20~mK.
The broad peak in $C(T)$ at~0.77~K is very likely to be associated with the formation of an extended short-range order, as it seems to coincide with the broad maxima in the $\chi(T)$ curves, below which a small difference between zero-field cooled data and field cooled susceptibility data has been reported~\cite{Hayes_2012}.
The observed temperature dependence of heat capacity $C(T) \sim T^2$ is not unusual in frustrated magnetism and has been seen previously, for example, in some Gd pyrochlores~\cite{Gd_pyros}. 

The magnetic entropy recovered by $T=5$~K is nearly 2R$\ln(2)$, which suggests that at least at low temperature SDO may be treated as an effective $s=1/2$ system.
This behaviour is also quite expected and commonly observed for Dy magnetic moments in the presence of a strong Ising-like magnetic anisotropy, for example in dysprosium pyrogermanate $\rm Dy_2Ge_2O_7$ \cite{Ke_2008} and dysprosium pyrochlore $\rm Dy_2Ti_2O_7$ \cite{Dy_pyros}.
In the case of SDO, high-temperature susceptibility data suggest that the $b$ axis is the easy direction for magnetisation~\cite{Hayes_2012}.

According to the inelastic neutron scattering data~\cite{CFE_Kenzelman}, the first well-defined crystal-field level for SDO is located at~4~meV (although a less pronounced peak at~2~meV, which is partially masked by the quasi-elastic signal, cannot be ruled out).
This observation implies that at approximately 20~K SDO will no longer behave as an effective spin 1/2 system.
 
From the data presented here (as well as from the previous magnetisation measurements~\cite{Hayes_2012}) it is obvious that
SDO behaves rather anisotropically in an applied field.
For $H \parallel [010]$, the field dependence of heat capacity measured at the base temperature (see Fig.~\ref{Fig3}) returns a small, but still well-defined peak at 2~kOe, as well as a much sharper and more pronounced double peak around 20~kOe.
Both the 2 and 20~kOe peaks correspond to sharp increases in the magnetisation (and therefore to peaks in $dM/dH$) shown in Fig.~5 of ref.~\cite{Hayes_2012}.
As the value of the magnetisation does not change significantly between these peaks and amounts to approximately a third of the total magnetisation for this direction of the applied field, it has been suggested~\cite{Hayes_2012} that the observed magnetisation plateau corresponds to the appearance of a collinear up-up-down ($uud$) structure, in which two thirds of magnetic moments are aligned along the field while the remaining third are aligned in the opposite direction.
The observed field dependence of the specific heat adds weight to this conjecture with the local maximum in $C(T,H)$ defining a region of stability of the collinear phase (see Fig.~\ref{Fig4}), which seems to extend at the lowest temperature from 2 to 19~kOe and to propagate to almost~1.5~K in 10~kOe.

At the lowest temperature the higher-field part of the phase diagram is separated from the lower-field part by a double phase transition, clearly seen in the $C(H)$ curves.
Remarkably, there is no indication in the magnetisation data~\cite{Hayes_2012} that the higher-field transition is actually split into two transitions and that the separation between them increases with increasing temperature.
After observing such a split from the heat capacity data we have made sure that this splitting is not caused by a slight misalignment of the sample or by the presence of several crystallographical domains, by verifying that two independently prepared and aligned samples of SDO return practically identical sets of specific heat data.

The sharp peaks in the $C(H)$ curves indicate multiple magnetic field-induced transitions in SDO for $H \parallel [010]$.
It is, however, impossible from the bulk-property measurements alone to state whether to not any of the field-induced phases are long-range in nature.
Careful neutron diffraction experiments are required to answer this question~\cite{further_neutrons}.

For $H \parallel [010]$, the field dependence of heat capacity differs significantly from the $H \parallel [001]$ case in that the latter does not show any signs of a field-induced phase transition.
A broad maximum in the lower-temperature $C(H)$ curves shown in Fig.~\ref{Fig5} corresponds to a similarly broad maximum in the $dM/dH$ curves for this direction of an applied field~\cite{Hayes_2012}.
In order to establish the reasons for such a dramatic difference in the magnetic behaviour of SDO for two different directions of an applied field, the crystal-field parameters must be established.
This exercise is far from trivial, as there are 8 Dy$^{3+}$ ions on the two district crystallographic sites in the unit cell (see Fig.~\ref{Fig1}).
The symmetry is rather low, therefore the number of crystal-field levels is expected to be large, but perhaps even more importantly, the positions of the levels at lower temperature can be influenced by the development of a short-range magnetic order.

The base temperature of the $^3$He cryostat used in our experiment has limited the lowest sample temperature to approximately 0.38~K.
In the context of quantum-critical transitions, it would be very interesting to extend the $C(H)$ measurements reported here to lower temperatures, particularly for $H \parallel [010]$ where the difference between the two field-induced transitions around 20~kOe seems to be decreasing with decreasing temperature.
However, this could prove to be experimentally challenging, as below 0.3~K (and extending down to at least 50~mK) there is a significant increase in the heat capacity of SDO~\cite{Hayes_2013}.
The origin of this upturn of the $C(T)$ curve below 0.3~K is presently unknown (it could potentially be caused by the hyperfine interactions); it results in a very rapid increase in $\tau$, the time constant of the measurements using the relaxation technique, even for very small samples, making the measurement times prohibitively long.

Another possible extension of the work presented here would be to measure the heat capacity of SDO for $H \parallel [100]$.
Given the difficulties associated with unfavourable sample geometry and also the featureless magnetisation curve for this direction of the applied field~\cite{Hayes_2012}, this research avenue does not seem to be particularly promising.

To summarise, we have investigated the low-temperature behaviour of \sdo\ in an applied magnetic field by measuring its specific heat for $H \parallel [010]$ and $H \parallel [001]$.
The collected data indicate that (i) in zero field, the material remains disordered down to 0.39~K (ii) for $H \parallel [010]$, a sequence of magnetically ordered phases is induced (iii) for $H \parallel [010]$, no transition to an ordered state occurs.
The results call for further neutron scattering experiments to clarify the nature of the field-induced phases of this geometrically frustrated compound.

\ack The authors acknowledge financial support from the EPSRC, UK under the grant EP/E011802/1.
Some of the equipment used in this research was obtained through the Science City Advanced Materials project: Creating and Characterising Next Generation Advanced Materials, with support from Advantage West Midlands (AWM) and was part funded by the European Regional Development Fund (ERDF).


\begin{thebibliography}{99}
\bibitem{caution} One has to be careful to distinguish the case where the absence of the LRO is due to frustration from the situation where the magnetic anisotropy is strong compared to the exchange interaction (so-called singlet-ground-state magnets). 
In the context of frustrated magnetism the singlet-ground state compounds CsFeCl$_3$ and CsFeBr$_3$ have been found to demonstrate a field-induced magnetic order.
For a review of their properties see Ref.~\cite{review}.
\end{thebibliography}
\end{document}